\documentclass[superscriptaddress, preprint, prb]{revtex4-2}
\usepackage{graphicx}
\newcommand{\NIMS}{National Institute for Materials Science, 1-2-1 Sengen, Tsukuba, Japan}
\newcommand{\UT}{University of Tsukuba, 1-1-1 Tennoidai, Tsukuba, Japan}

\begin{document}
\title{Neural networks for a quick access to a digital twin of scanning physical properties measurements}% Force line breaks with \\

\author{Kensei Terashima}
\email{TERASHIMA.Kensei@nims.go.jp}
\affiliation{\NIMS}
\author{Pedro Baptista de Castro}
\affiliation{\NIMS}
\affiliation{\UT}
\author{Miren Garbi\~{n}e Esparza Echevarria}
\affiliation{\NIMS}
\affiliation{\UT}
\author{Ryo Matsumoto}
\affiliation{\NIMS}
\author{Takafumi D Yamamoto}
\affiliation{\NIMS}
\author{Akiko T Saito}
\affiliation{\NIMS}
\author{Hiroyuki Takeya}
\affiliation{\NIMS}
\author{Yoshihiko Takano}
\affiliation{\NIMS}
\affiliation{\UT}

% Keywords: Please provide a minimum of three and a maximum of seven keywords, separated by commas

% Abstract should be written in the present tense and impersonal style (i.e., avoid we), and be at most 200 words long
\begin{abstract}
  For performing successful measurements within limited experimental time, efficient use of preliminary data plays a crucial role.
  This work shows that a simple feedforward type neural networks approach for learning preliminary experimental data can provide quick access to simulate the experiment within the learned range. The approach is especially beneficial for physical properties measurements with scanning on multiple axes, where derivative or integration of data are required to obtain the objective quantity. Due to its simplicity, the learning process is fast enough for the users to perform learning and simulation on-the-fly by using a combination of open-source optimization techniques and deep-learning libraries. Here such a tool for augmenting the experimental data is proposed, aiming to help researchers to decide the most suitable experimental conditions before performing costly experiments in real.  Furthermore, this tool can also be used from the perspective of taking advantage of reutilizing and repurposing previously published data, accelerating data-driven exploration of functional materials.
\end{abstract}
\maketitle
% Text: Please use section headings and subheadings as specified below. For communications, all section headings apart from Experimental Section should be removed
% Please make the first reference to a display item bold: \textbf{Figure 1}
% Do not abbreviate Figure, Equation, etc.; display items are always singular, i.e., Figure 1 and 2.
% Equations are always singular, i.e., Equation 1 and 2, and should be inserted using the {equation} environment, not as graphics
% Please do not use footnotes in the text, additional information can be added to the Reference list.

\section{Introduction}
The recent global rise of databases on materials as well as accessible repositories and advanced text mining techniques has led to the evolution of materials informatics as an effective tool for exploration and design of functional materials \cite{Stanev_2021ComMat}.   Along with the increase of the available crystal structural input data, the emergence of high-throughput first-principles calculation systems has enhanced a variety of materials estimated property data useful for statistical analysis \cite{materialsprojectJain2013, oqmdkirklin2015open, aflowcurtarolo2012aflow}.  In addition, the application of machine-learning methods has proved to be quite efficient for dealing with a large amount of materials data \cite{Stanev_2021ComMat,Castro2020-12}, allowing the extraction of useful information that has been hidden or too complicated for an ordinal human to percept.  As a result of the recent rapid growth of such tools for handling materials data, the process for obtaining candidate materials in cyberspace has been significantly accelerated, while the bottleneck for the discovery of functional materials remains on the actual synthesis and experimental evaluation in real space \cite{xiong2021optimizing}.  This is because the cost for obtaining experimental data is usually high, especially when researchers have to use a state-of-the-art shared facility with fixed machine time. Furthermore, not only the experiment itself but also its preparation can be time-consuming as sometimes one has to start from optimization of the sample synthesis process, a typical aspect for physics and materials science measurements. \\

Such a costly-in-real situation is even more serious in industrial fields, where a large amount of money has to be spent for the production of "real" items.  To overcome this issue, there an idea called ``digital twin" was produced, where one builds a simulation model that takes account of the data in real space and tries to optimize the parameters in the cyber space before real synthesis \cite{Tao_2019DT}.  For being beneficial, the model has to correspond enough to the objective, and the cost for the simulation both in money and time has to be lower than the actual production.  Here again, the application of machine-learning methods has greatly reduced the duration required for the execution of simulation as compared to that for performing conventional simulation algorithms.  Recently, the use cases of such a digital-twin approach aided by machine learning are becoming popular also in academic research. For instance, machine-learning was used for approximating time-consuming simulation results for fluid dynamics, allowing researchers to perform a quick optimization of the growing condition of objective crystals \cite{Dang_2021DT}.\\

In this context, such a digital-twin approach is expected to be effective and highly useful also for physics and materials science measurements, where the objective properties are sometimes deduced from experimental data by applying mathematical operations such as derivatives or integration.  Especially in physics, the development of new approaches or ideas to data analysis, creates an opportunity to revisit or reanalyze previously published data, allowing researchers to draw new insights from this previous knowledge.  However, it is quite often that the data are non-evenly spaced along multiple axes due to experimental confinements, and it may contain non-linear responses that cannot be interpolated easily.  In such a case, it is quite difficult to estimate the target quantity from preliminary data.  On the other hand, artificial neural networks are known to have high flexibility and adaptivity in principle \cite{Cybenko_1989, Mhaskar_Miccelli_1992}.  It has been also discussed \cite{Mhaskar_1996} that neural networks learning can be as efficient as other approximation methods from a viewpoint of minimizing loss using the number of parameters as low as possible. They have proved to be even useful when the objective has inhomogeneity in smoothness \cite{Suzuki_2018}, or when the objective has several discontinuities \cite{pmlr-v89-imaizumi19a}. Thus, approximation using this method is expected to be suitable for materials science and physics experimental data that tend to contain discrete change due to first-order transition and/or multiple peak structures.

In this article, we show that learning preliminary data of materials science by neural networks provides us a way to quickly build a model that can describe the target property dependencies on scanning axes (treated as features). In order to ensure the accuracy of the model that can be built on-the-fly during the experiment, several open python libraries have been used in combination, as well as several tools for efficient learning. Once the model is built, the simulation within the learned feature range can be done instantly, which enables researchers to evaluate the experimental plan as well as its cost (\textbf{Figure 1(a)}).  We propose that such an approach of data augmentation helps researchers to use preliminary- or deposited data effectively for performing the data-driven search of functional materials.\\

\begin{figure}[!htpb]
  \includegraphics[width=\textwidth]{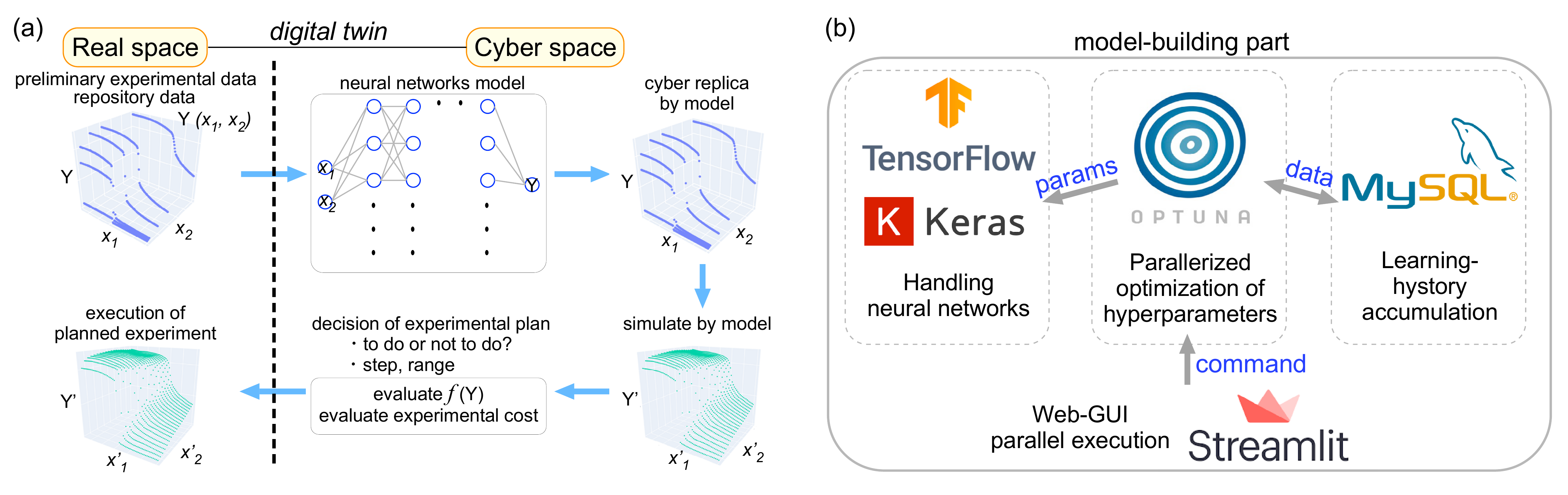}
  \caption{(a) Graphical concept for application of digital-twin approach to materials property measurements for performing efficient experiments using preliminary data as train data. (b) Combination of open libraries used in this study to quickly build a fully-connected feedforward neural networks model for simulating experimental data.}
  \label{fig-concept}
\end{figure}

\section{Construction of neural networks model}
In our approach, the constructed code relies mainly on 4 open libraries available to the public in addition to the Anaconda package \cite{anaconda}, shown in \textbf{Figure 1(b)}. As a machine-learning body, we built fully-connected feedforward neural networks, using Keras with Tensorflow backend (v.2.x) \cite{chollet2015keras, tensorflow2015-whitepaper}.  Keras has several optimizers by default, and among them, we tried stochastic gradient descent, Nesterov accelerated gradient \cite{Nesterov_1983}, root mean square propagation \cite{Tieleman_Hinton_2012}, and adaptive moment estimation (Adam) \cite{KingmaB14} for several times to our data shown in \textbf{Section 3}. For our data, Adam tended to be the most stable and fast for minimizing loss of mean squared error between predicted values and training data (not shown). We set the number of layers, nodes, learning rate, and batch size as hyperparameters to be optimized. Once those hyperparameters are suggested, random weights and biases are given as initial values in Keras and they are tuned by the optimizer as the epoch proceeds in a single training run. Thus the learned result in a single run varies by the initial random values, even though there could be a general trend for obtaining a moderate set of hyperparameters for example the case in learning rate and batch size \cite{Krizhevsky_2014, Goyal_2017, JastrzebskiKABF18}. To be quick, we confined the search space of the hyperparameters in the following range: $2\leq$ number of layers $\leq10$, $50\leq$ number of nodes $\leq200$, $1e^{-5}\leq$ learning rate $\leq1e^{-4}$, and $16\leq$ batch size $\leq1024$ varying on the data size.  To obtain the best possible learned result in a fixed time, we used the Bayesian optimization package Optuna \cite{optuna_2019}. By combining a relational database provided by MySQL \cite{widenius2002mysql}, Optuna receives each training result with a set of hyperparameters running in parallel, so that it can determine a set of hyperparameters for the next run based on Tree Parzen Estimator \cite{Bergstra2015}.  Typically, the best set of hyperparameters is found by at most $\sim$30 runs.  After many trials, we found that overfitting is welcome for our purpose to learn the detail of the real experimental data (including noise level and characteristics of experimental machines), and that cross-validation simply increases the required time for training. Thus we used the whole data for training.  For visualization of data and controlling multiple Optuna runs in parallel, we used the Streamlit library \cite{streamlit} to construct a graphical user interface (GUI) that works on the web browser. \\

To obtain the best model during the hyperparameter search, we used three functions equipped in the above libraries.  First, is early stopping implemented in Keras, with the option of saving the best model in the run, where the maximum epoch number is 500.  The second is the learning rate scheduler implemented in Keras, where we change the learning rate to be 10 times smaller value from its initial value after 100 epochs. The third is pruning with the Asynchronous Successive Halving algorithm \cite{ASHA} implemented in Optuna, where it examines if each run shall be pruned based on the comparison of current score with those of running parallel by 25$^n$ epochs where $n$ is an integer greater than or equal to zero. \\

The approach used here might be primitive as compared to recent state-of-the-art combinations of neural networks architectures, optimizers, and other techniques, but we tried to keep the system simple, fast, and easy to use as much as possible, considering that the target users are not specialist of data science but materials scientists like us. As a result, we could quickly finish learning small datasets typical for physical property measurements (number of data $\sim$ a few thousand) shown in the next section, even without using graphics processing units (GPUs). More detail on case studies for the duration and used computers are in Supporting Information. A tutorial for how to get ready and perform the neural networks learning and simulation shown in this paper will be available at https://www.github.com/kensei-te/mat$\_$interp.  \\

\section{Results and Discussion}
In this section, we show 4 typical use cases as an example.  The first is the magnetization curve of a well-known magnetocaloric material ErCo$_{2}$, where $1^{st}$ order transition occurs that accompanies a steep change in the magnetization as a function of temperature and the transition temperature migrates by application of an external magnetic field.  The second example is the magnetization curve and estimation of magnetocaloric effect in another magnet Fe$_{3}$Ga$_{4}$ that shows $1^{st}$ order transition but its transition temperature changes more drastically by applied field than in the first example of ErCo$_{2}$.  The third example is the resistance data of the PdH superconductor which is relatively noisy as compared to above examples.  The fourth example is a mapping data of angle-resolved photoelectron spectroscopy (ARPES) intensity on Fermi energy of La(O,F)BiS$_{2}$ superconductor taken above superconducting temperature ($T_c$) as a demonstration of the application of the method for two-dimensional intensity map, consisting of a number of peak structures.

\subsection{$M(T,H)$ of ErCo$_2$ magnet and estimation of its magnetocaloric effect}\label{section-ErCo2}
First, we show a case for physical property data that have 1$^{st}$ order transition, namely a steep change in the observed value. ErCo$_2$ is known to be a ferrimagnet with its ordering temperature $T_{Curie}$ of $\sim$ 35 K \cite{Ross_PhysRev_1964}. It exhibits a $1^{st}$ order magnetic transition accompanied by magnetostriction, and the material is one of the most popular materials for magnetic refrigeration \cite{Wada,GschneidnerJr_2005,Amin_NCommun_2022}.  To evaluate the magnitude of magnetic entropy change $|\Delta S_{M}|$, one of the most common ways \cite{GschneidnerJr_2005} is to measure the magnetization of the sample as a function of temperature and applied magnetic field change $\Delta H$(from 0 to $H$) either isothermal or isofield, and deduce $|\Delta S_{M}|$ by applying following Maxwell's equation for the observed magnetization data:\\  
\begin{displaymath}
  \Delta S_{M}(T,\Delta H) =  \mu_0 \int_{0}^{H} \frac{\partial M(T,H^{'})}{\partial T} dH^{'}
\end{displaymath}

The equation includes an integration along the applied field ($H$).  Practically, with such discrete experimental data, trapezoidal integration can be applied for a numerical integration, which corresponds to a linear interpolation along the $H$-axis.  However, if we estimate $|\Delta S_{M}|$ in this manner from the experimental data shown in \textbf{Figure 2(a)} taken by a coarse magnetic field step, it ends up with extrinsic oscillation of estimated $|\Delta S_{M}|$ values as shown as a gray solid line in \textbf{Figure 2(e)}.  This is simply because of the failure of linear interpolation for data having $1^{st}$ order transition (see Supporting Information for technical details).  To avoid this extrinsic effect for estimation of the target property, one has to use a finer magnetic field step, so that linear interpolation starts to work well since $\partial M/\partial T|{_H}$ between two steps has a finite overlap each other.  This means one has to know the proper measurement step before the measurement itself.  In such a case, neural networks simulation based on preliminary data works efficiently as it can perform a non-linear interpolation.  \textbf{Figure 2(b)} shows the predicted magnetization values by a constructed model for the conditions in the train data of \textbf{Figure 2(a)}, which corresponds well to the train data, indicating that a plausible model has been made. Once such a model is constructed, the model can perform quick predictions for any desired feature values within the learned range.  After a couple of trials, we found in \textbf{Figure 2(f)} that extrinsic oscillation in the simulation is expected to be significantly suppressed when the magnetic field step of the measurement is smaller than 0.2 T.  We measured the magnetization in the proposed step as in \textbf{Figure 2(c)}, and found in \textbf{Figure 2(e)} that the experimentally evaluated $|\Delta S_{M}|$ corresponds very well to what has been predicted by the simulated model.

% \begin{itemize}
%   \item 
%   \item 
%   \item 
%   \item 
% \end{itemize}

\begin{figure}[!htpb]
  \centering
  \includegraphics[width=90mm]{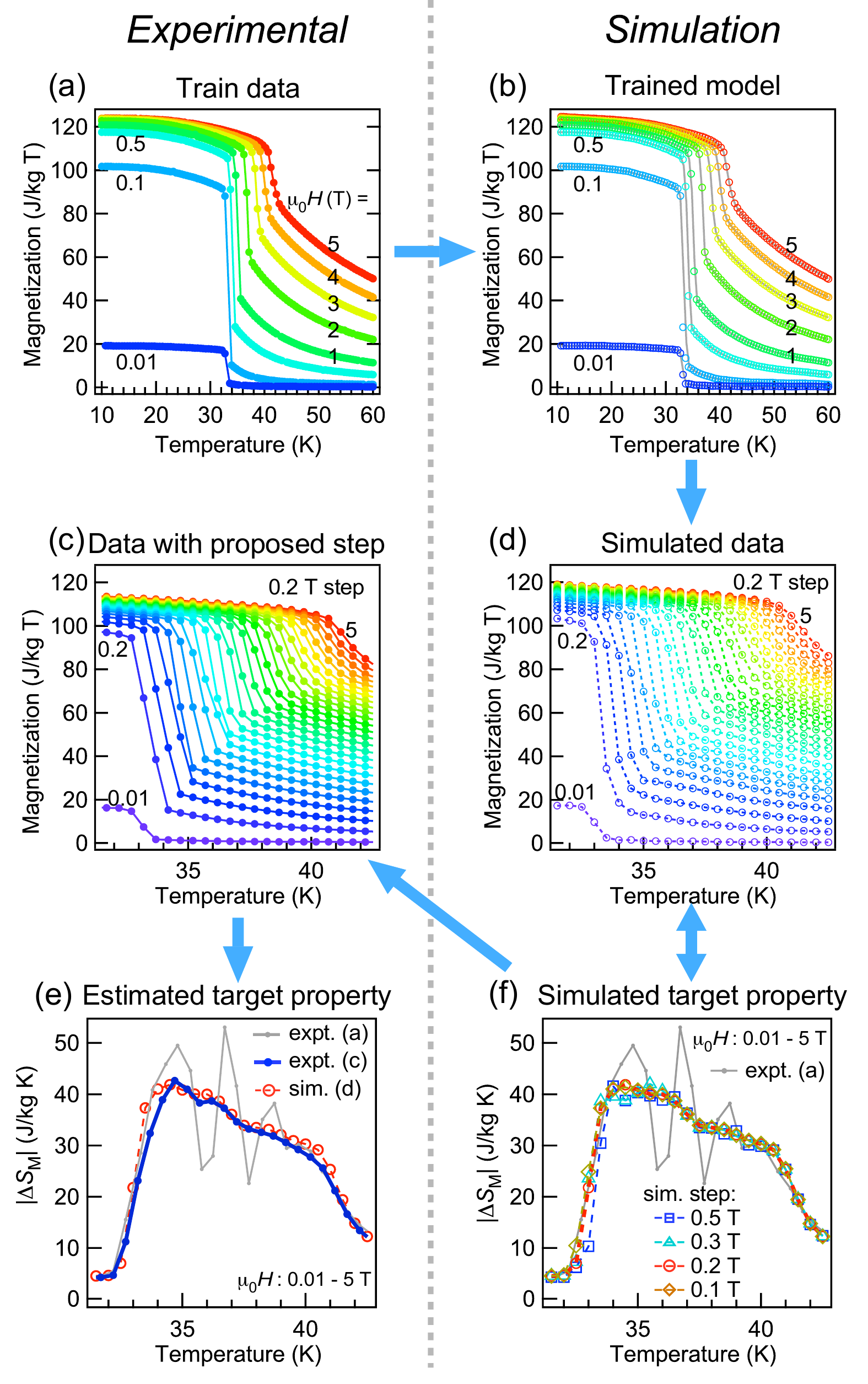}
  \caption{(a) Preliminary magnetization data of ErCo$_2$, which is used as the train data for constuction of model. (b) Simulated magnetization by the trained model for the conditions in the train data. Gray solid lines show the train data. (c) Magnetization of ErCo$_2$ with magnetic field step proposed by the model. (d) Simulated magnetization by the trained model.  (e) Estimated $|\Delta S_{M}|$ of ErCo$_2$ for a field change of 0-5 T. (f) Simulated $|\Delta S_{M}|$ for different experimental steps of magnetic fields. Gray solid line shows the estimated $|\Delta S_{M}|$ from (a) by applying trapezoidal integration along magnetic field direction. Blue arrows between figures indicate an order of workflow of the measurements and simulations.}
  \label{fig-section3_1}
\end{figure}

\subsection{$M(T,H)$ of Fe$_3$Ga$_4$ magnet and estimation of its magnetocaloric effect}\label{section-Fe3Ga4}
Here we show the case of another magnet Fe$_3$Ga$_4$ and evaluation of its magnetocaloric property, to show the possibility of applying neural networks study for analyzing literature data.  The magnetic property and magnetization as a function of temperature and magnetic field have been reported \cite{Kawamiya_Fe3Ga4_JPSJ1986}, while $|\Delta S_{M}|$ of this material has not been reported yet. \textbf{Figure 3(a)} shows our preliminary experimental data that corresponds well to the literature data \cite{Kawamiya_Fe3Ga4_JPSJ1986}, but both of them do not have fine enough experimental steps to evaluate $|\Delta S_{M}|$ of this material.  As shown in \textbf{Figure 3(d)} and \textbf{3(f)}, we have performed a simulation of expected $|\Delta S_{M}|$ taken on each field step case and found that extrinsic effect for $|\Delta S_{M}|$ would not be observed when we measure magnetization by a magnetic field step less than 0.1 T for this material.  Considering the required experimental time, we chose a 0.075 T step, and the simulated data showed an excellent correspondence with the real experimental data as shown in \textbf{Figure 3(c)-(f)}.  The simulation also tells us that, the estimated $|\Delta S_{M}|$ is at most 0.25 J kg$^{-1}$ K$^{-1}$ by a field change of 0-1 T (which is small as compared to those of other magnetocaloric materials \cite{GschneidnerJr_2005}) despite this measurement covers wide temperature range of 10-200 K with fine magnetic field step, hence it is expected to take approximately 1 week.  Therefore if one is simply looking for a material with high $|\Delta S_{M}|$, the simulation tells us that this material may not be suitable to be synthesized and measured by spending a fair amount of time. In other words, such a simulation is not only useful for analyzing the past preliminary or repository data from a different point of view, but also it can help researchers to evaluate the cost of performing the experiment by leveraging the confined knowledge available from preliminary- or literature data.
 
\begin{figure}[!htpb]
  \centering
  \includegraphics[width=90mm]{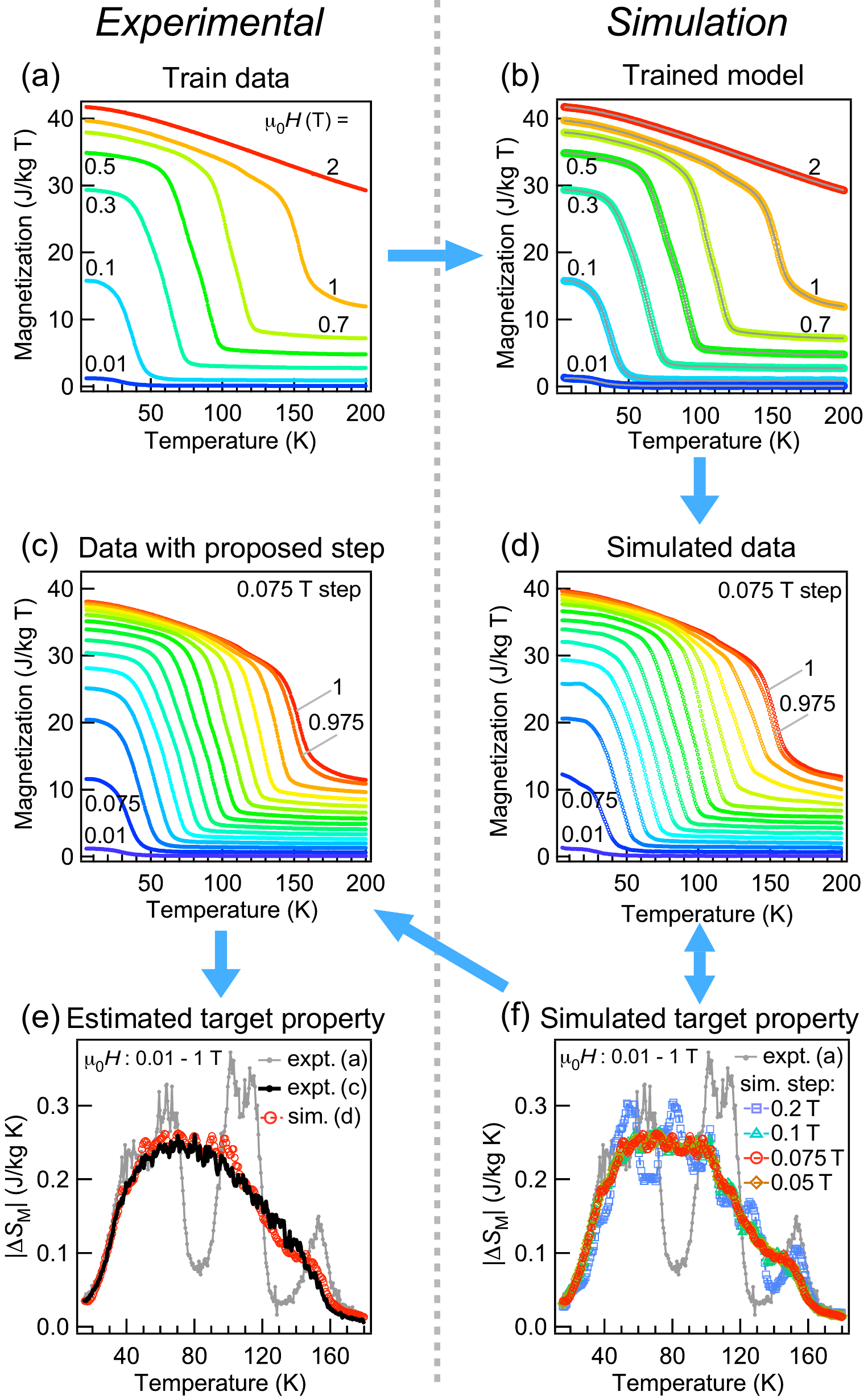}
  \caption{(a) Preliminary magnetization data of Fe$_3$Ga$_4$, which is used as the train data for constuction of model. (b) Simulated magnetization by the trained model for the conditions in the train data. Gray solid lines show the train data. (c) Magnetization of Fe$_3$Ga$_4$ with magnetic field step proposed by the model. (d) Simulated magnetization by the trained model.  (e) Estimated $|\Delta S_{M}|$ of Fe$_3$Ga$_4$ for a field change of 0-1 T. (f) Simulated $|\Delta S_{M}|$ for different experimental steps of magnetic fields. Gray solid line shows the estimated $|\Delta S_{M}|$ from (a) by applying trapezoidal integration along magnetic field direction. Blue arrows between figures indicate an order of workflow of the measurements and simulations.}
  \label{fig-section3_2}
\end{figure}

\subsection{$R(T,H)$ of PdH superconductor}\label{section-rho-T}

Next, we show the case for data with a lower signal-to-noise ratio. For this purpose, we used a part of our resistance data of PdH \cite{PdH} as train data (\textbf{Figure 4(a)}).  As can be seen in \textbf{Figure 4(b)}, the model does not necessarily capture the ultrafine structure coming from noise, however it grabs not only the overall tendency of the lineshape but also several fine structures that might be influenced by noise.  The inability of tracking the full noise is because of the fact that the number of layers and nodes in the model is limited and lower than the number of total train data points.  Indeed, neural networks are sometimes used for denoising purposes \cite{Kim_denoise_RSI2021,Restrepo_denoise}.  On the other hand, our intention here is to simulate the experiment as it is, including the extrinsic effect such as noise and unique characteristics of the measurement apparatus, based on the non-evenly spaced preliminary data.  Comparing the experimental data in \textbf{Figure 4(c)} with simulated data in \textbf{Figure 4(d)}, it can be said that the model in the current approach can tell us how the expected lineshapes can be affected by the noise level appeared in the train data, with a finite inevitable denoising effect.\\

\begin{figure}[!htpb]
  \centering
  \includegraphics[width=90mm]{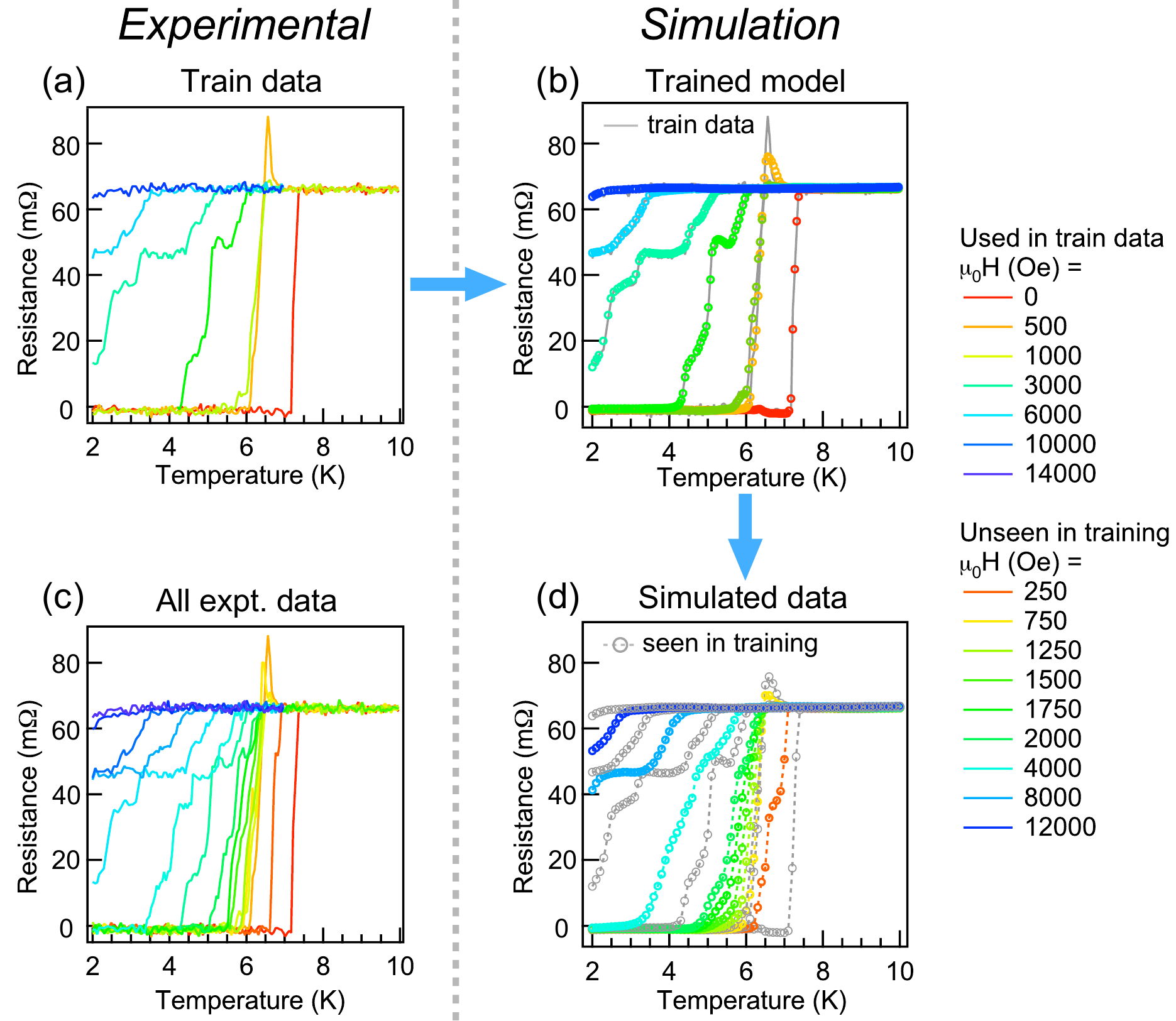}
  \caption{(a) Train data of resistivity measurements for PdH, taken from a part of the whole experimental data in (c). (b) Simulated resistance by the trained model for the conditions in the train data. Gray solid lines show the train data. (c) Experimental data of resistivity measurements for PdH superconductor \cite{PdH}. (d) Simulated resistance by the trained model for the same conditions in (c). Gray open circles show the conditions seen by the model during training.}
  \label{fig-section3_3}
\end{figure} 

\subsection{Angle-resolved photoelectron spectroscopy intensity map of La(O,F)BiS$_2$}\label{section-ARPES}

As a last example, we show that our quick and simple approach is also applicable to two-dimensional intensity map, namely, ARPES intensity map at Fermi energy in La(O,F)BiS$_2$ superconductor on normal state (above $T_c$).  ARPES intensity at fixed energy tends to be high in angular area where corresponding electronic state exists, thus the data consists of a number of peak structures (see Supporting Information for lineshape of each cut).  In \textbf{Figure 5}, the ARPES intensity is shown as a function of two-dimentional angles with respect to the normal angle to the sample surface, that corresponds to two-dimentional wave vectors in reciprocal space \cite{Damascelli}. For constructing such a mapping data, it is quite often that one measurement is performed with fixing one of the axis (for instance $\theta_2$ in the figure) and keep concatenating data with different angle of $\theta_2$. Therefore a preliminary measurement with coarse angle step of $\theta_2$ (\textbf{Figure 5(a)}) can be performed before a serious measurement with fine angular steps and better statistics (\textbf{Figure 5(c)}), so that one can determine the objective measurement area of angles.  With help of neural networks learning, we can simulate how data would become with specific measurement area and steps as in \textbf{Figure 5(d)}, thus the simulation helps researchers to make a decision for planning the experiment. Here we note that the simulated pattern captures characteristics of the measurement apparatus, namely high sensitivity at edge and detection limit in certain angle, so it is suitable for estimating the effect of such instrumental conditions as well. However, the simulated pattern is affected by the presence of sizable noise, hence the pre-processing of train data with other noise reduction method might help further though such process may take longer time.  Alternatively, it is also possible to take an avarage of predictions by several models found during learning process.  We also note that although our simple approach is applicable on-the-fly to such a mapping data including non-evenly spaced ones, other sophisticated methods such as super resolution convolutional neural networks-based approach \cite{Kim_2016_SR_deepCNN, Chai_SciRep_2020, Peng_RSI_2020} would perform certainly better especially for evenly spaced data such as photographs if one can spare enough time for training and has an access to powerful GPUs.

\begin{figure}[!htpb]
  \centering
  \includegraphics[width=90mm]{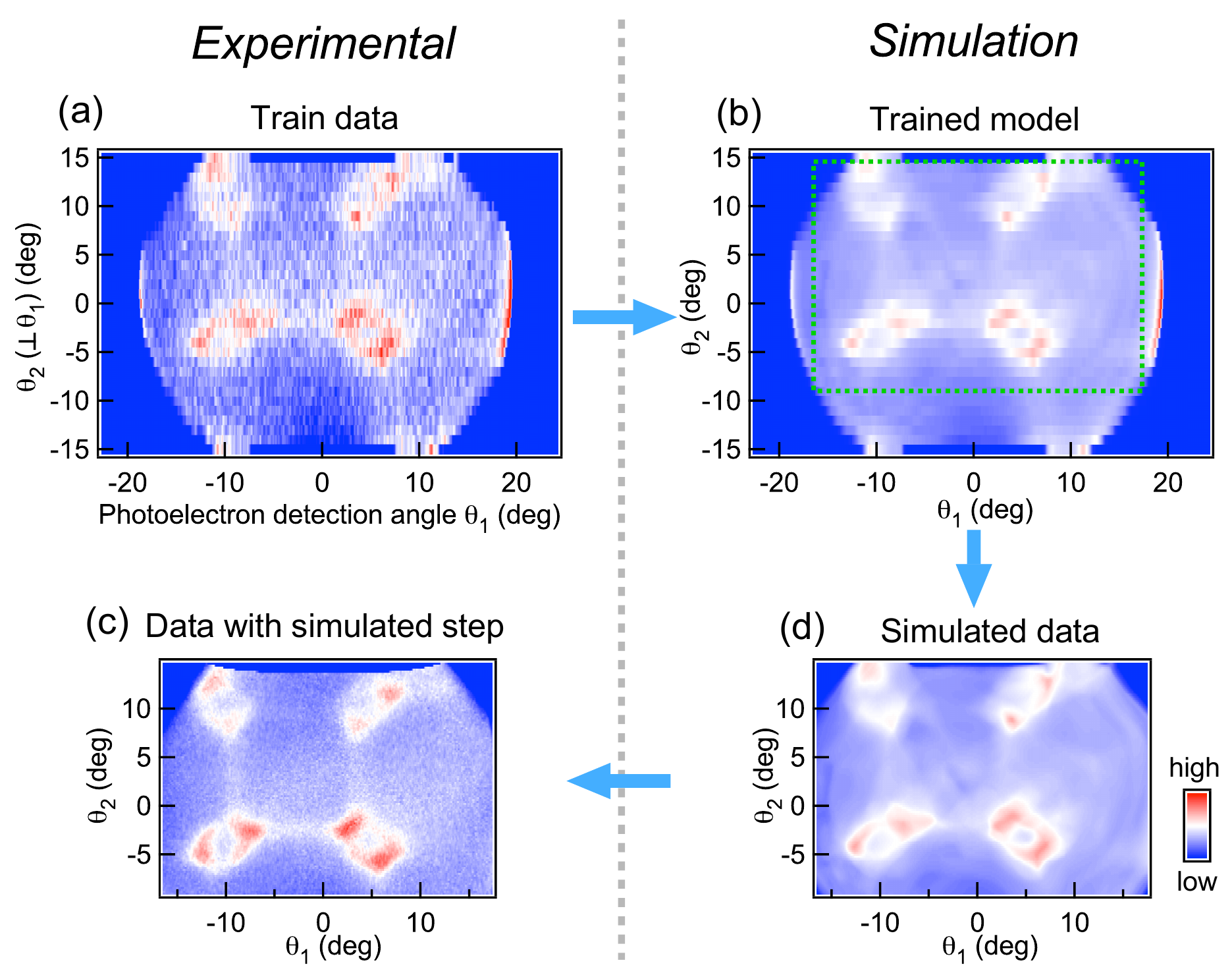}
  \caption{(a) Roughly-scanned ARPES intensity distribution as a function of angles with respect to sample normal, used as train data. (b) Simulated ARPES intensity map by the trained model for the conditions in the train data. Green dotted square shows the simulated angular area in (d). (c) Same as (a) but with proposed angular steps by simulation in (d). (d) Simulated ARPES intensity map by the trained model for estimation of measurement conditions in (c).}
  \label{fig-section3_4}
\end{figure}

\subsection{Scope and limitation of current approach}\label{section-scope}
Here we note the applicability/inability of our approach.  The code is available at \\https://www.github.com/kensei-te/mat$\_$interp including how to get ready to use GUI, which runs in Linux and Mac OS (see Supporting Information for details).  It can learn quickly the data so that one can perform the simulation during the experiment, especially when the total number of data points kept up to a few tens of thousands. For this reason, it is preferred to keep the dimension of data (i.e. number of feature columns) low, even though it can deal with multidimensional data.   The approach would be useful mainly for two cases:  (i) Simulation of a costly experiment from preliminary experimental data: based on the roughly-scanned data, one can simulate how the data and resultant estimated target property would be, when taken in fixed experimental steps and ranges.  Thus it helps researchers to plan and perform efficient experiments.  This is beneficial especially when the target properties are deduced from mathematical operations of experimental data and hence are depending on experimental conditions, as shown in \textbf{Section 3.1} and \textbf{3.2}.  (ii) Aligning and augmentation of repository data: there is an increasing number of accessible materials data sets, that have been published in the past. It is quite often that one encounters such problems that despite the material being of interest, there is a mismatch or lack in the experimental conditions as compared to what is required in order to perform additional analysis. As the current method can adapt readily to non-evenly spaced data with a non-linear response, it will help researchers to judge whether the material is worthy to be examined further by reusing this data for the simulation method proposed here, to spare a certain amount of experimental cost as shown in \textbf{Section 3.2}. \\

On the other hand, the method is ultimately a mere interpolation, thus the signal that is skipped in the train data with too coarse step will never be predicted in the simulation. For the same reason, though the simulation helps researchers to plan the experiment, the simulated results and estimated values of physical properties themselves should be clearly distinguished from experimental data and be taken with care since it is not certain what will come out in reality until verified by experiment. Also, the prediction by learned models includes how the data will be affected by unique characteristics of experimental apparatus (sensitivity, detection limit, and so on) and noise, as has been shown in \textbf{Section 3.3} and \textbf{3.4}.  The effect of noise could be examined by comparing several simulated results in different runs (see Supporting Information for details).  If the researcher aims to suppress the effect of noise rather than simulating it, other noise reduction methods may work better.\\

\section{Conclusion}
In this work, we present a simple and quick method to simulate experimental data by learning preliminary data using fully-connected feedforward neural networks.  The approach is shown to be suitable for materials science experimental data with typical examples.  Such a tool would help researchers to deal with preliminary- or past repository data efficiently, supporting decision-making during the experimental research. A tutorial on how to get ready and perform the neural networks learning and simulation shown in this paper will be available at https://www.github.com/kensei-te/mat$\_$interp.

\section{Experimental Section}\label{exp-sec}

\subsection{Sample Synthesis and Preparation}
ErCo$_2$ sample was obtained by arc-melting of Er and Co rocks in a vacuum chamber of Ar atomosphere, followed by a 21 hour annealing on 850 \(^\circ\)C and quenching by water. 
Fe$_3$Ga$_4$ sample was obtained by arc-melting of Fe and Ga chunks. The synthesis procedure of PdH sample is provided in \cite{PdH}, while the synthesis procedure of La(O,F)BiS$_2$ sample is in \cite{Nagao}.
\\

\subsection{Experimental Conditions}
Magnetization of samples has been measured using Magnetic Properties Measurement System (Quantum Design), in field cooling process. Resistance of sample under 2.6 GPa has been measured by four probe method using diamond anvil cell with boron-doped diamond electrodes \cite{PdH} and Physical Properties Measurement System (Quantum Design). ARPES measurements have been performed at BL28A \cite{KitamuraRSIKEK} in Photon Factory KEK, Japan (proposal number 2021G686), with DA30 analyzer (Scienta Omicron) at 20 K (estimated $T_c$ of the sample is approximately 2.2 K \cite{SYamamoto}), $h\nu$ = 70 eV.

\section{Acknowledgements} 
The ARPES measurements were performed at BL28A of Photon Factory, KEK (proposal No. 2021G686). The authors acknowledge Wei-Sheng Wang and Taku Tou for testing the installation of the code. This work was supported by the JST-Mirai Program (Grant No. JPMJMI18A3), JSPS Bilateral Program (JPJSBP120214602), and JSPS KAKENHI (Grant Nos. 20K05070, 19H02177). P.B.C. and M.G.E.E. acknowledges the scholarship support from the Ministry of Education, Culture, Sports, Science and Technology (MEXT), Japan. 

% \bibliography{references}

% \includepdf{SI.pdf}
% \bibliographystyle{apsrev4-2}
\bibliography{references}

\end{document}

% Use the \preprint command to place your local institutional report
% number in the upper righthand corner of the title page in preprint mode.
% Multiple \preprint commands are allowed.
% Use the 'preprintnumbers' class option to override journal defaults
% to display numbers if necessary
%\preprint{}

\title{Supporting Information for \\ Neural networks for a quick access to a digital twin of scanning physical properties measurements}% Force line breaks with \\

\author{Kensei Terashima}
\email{TERASHIMA.Kensei@nims.go.jp}
\affiliation{\NIMS}
\author{Pedro Baptista de Castro}
\affiliation{\NIMS}
\affiliation{\UT}
\author{Miren Garbi\~{n}e Esparza Echevarria}
\affiliation{\NIMS}
\affiliation{\UT}
\author{Ryo Matsumoto}
\affiliation{\NIMS}
\author{Takafumi D Yamamoto}
\affiliation{\NIMS}
\author{Akiko T Saito}
\affiliation{\NIMS}
\author{Hiroyuki Takeya}
\affiliation{\NIMS}
\author{Yoshihiko Takano}
\affiliation{\NIMS}
\affiliation{\UT}

%\date{\today}% It is always \today, today,
             %  but any date may be explicitly specified

%\keywords{Suggested keywords}%Use showkeys class option if keyword
                              %display desired
\maketitle

% \tableofcontents
% Trapezoidal integration and bilinear interporation  
% Operating system and installation guide  
% Example of duration to perform neural networks learning  
% Stability of learned and simulated results
% Spectral lineshapes of ARPES image plot (Section3.4)

% body of paper here - Use proper section commands
% References should be done using the \cite, \ref, and \label commands
\newpage
\section{Trapezoidal integration and bilinear interporation}

Trapezoidal integration is a numerical method to often used to obtain the an approximation of the area of interest below the curve of a function $f(x)$ and between the a given set of desired points $a$ and $b$, assuming: $\int_{a}^{b} f(x) dx \sim \frac{(b-a)}{2} \left(f(a)+f(b)\right)$, 
that corresponds to a summation of linearly-interpolated values of $f(x)$ between starting point $a$ and end point $b$.  Sometimes, for some datasets this linear interpolation method can fail. Figure SM1 shows a typical example of such a failure in linear interpolation.  Suppose there are data for $y(x_{1}, x_{2})$ taken in $x_2$ = $a$ and $b$, where peak position of $y$ migrates by changing $x_2$ value, as in Figure SM1(a). If one performs bilinear interpolation as in Figure SM1(b), it does not succeed to reproduce what is desired (as in Figure SM1(a)) but it produces an artificial dip in the obtained interpolated $y(x^{i}_{1},x^{i}_{2})$ .In Section 3.1 of the manuscript, $x_{1}, x_{2}, y$ correspond to temperature, magnetic field, and $\left(\frac{\partial M}{\partial T}\right)$, respectively, namely $y(x_{1}, x_{2}) = \frac{\partial M(T,H)}{\partial T}$.  This causes extrinsic oscillation in $|\Delta S_{M}|$ derived by integrating $\left(\frac{\partial M}{\partial T}\right)$ along $H$ direction.

\begin{figure}[h]
  \includegraphics[width=15 cm]{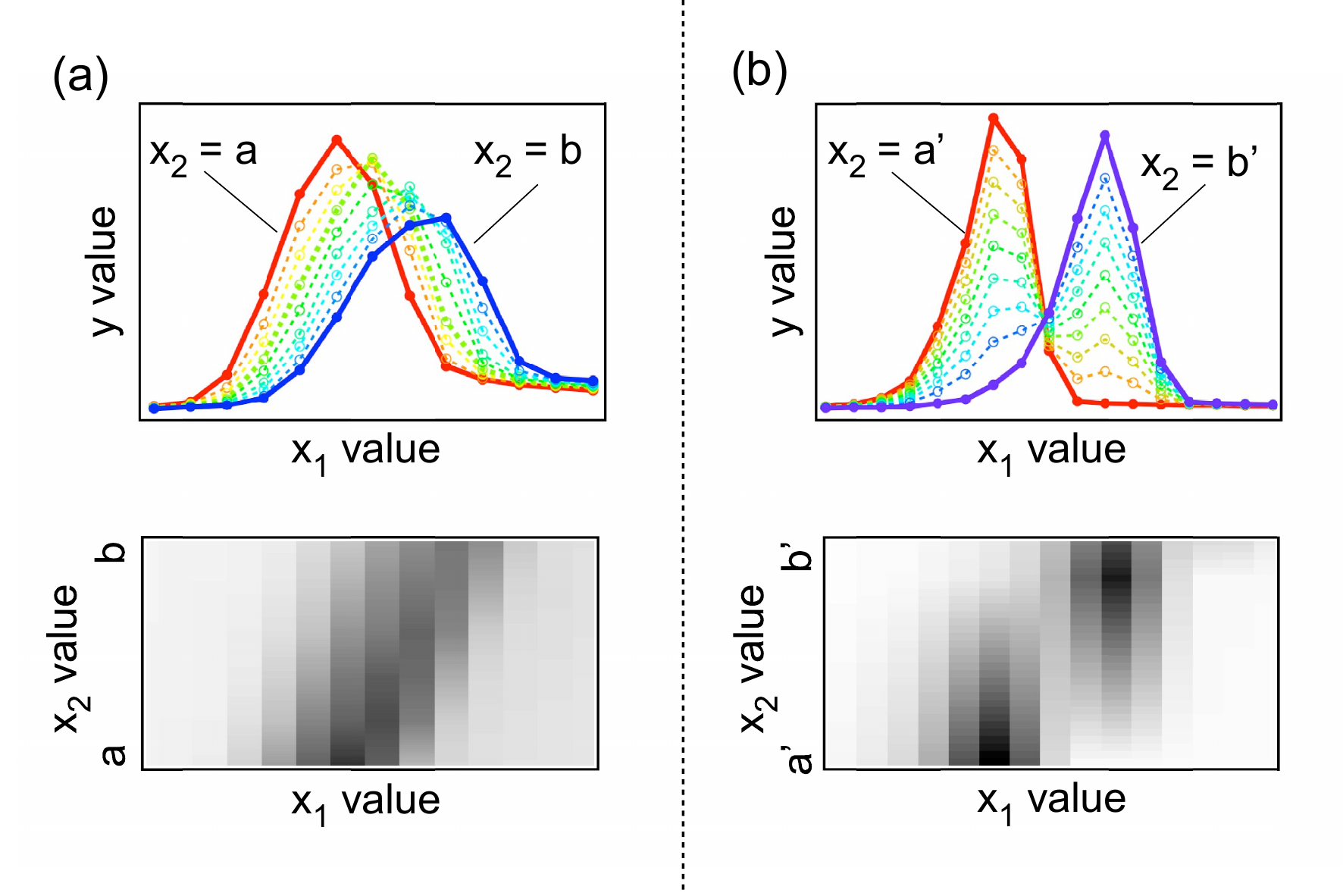}
  \caption{Schematic graph showing how (bi)linear interpolation fails to grab characteristic change. Top graphs show spectral lineshapes and bottom graphs are corresponding intensity plots. (a) What is expected (b) What (bi)linear interpolation gives.}
  \label{fig:linear}
\end{figure}

We also note here that the situation is also valid for isothermal cuts for $M(T,H)$.  In case of isothermal cuts, $|\Delta$$S_{M}|$ can be deduced by $\Delta S_{M}(T_{i},H) = \Sigma_{i} \frac{M_{i+1}-M_{i}}{T_{i+1}-T_{i}} \Delta H_{i}$ (in case of using forward differential), which corresponds to the gray shaded area as depicted in Figure SM2(a).  As can be seen in Figure SM2(a) in case of coarse step of $H$, linear interpolation along $H$-direction (corresponds to dotted line in the Figure) creates an extrinsic oscillation of area between two subsequent measured temperatures.  Figure SM2(b) shows the simulated $M(T,H)$ by neural networks model built by Figure SM2(a) train data, where such extrinsic oscillation of area between measured temperatures is suppressed.

\begin{figure}[h]
  \includegraphics[width=14 cm]{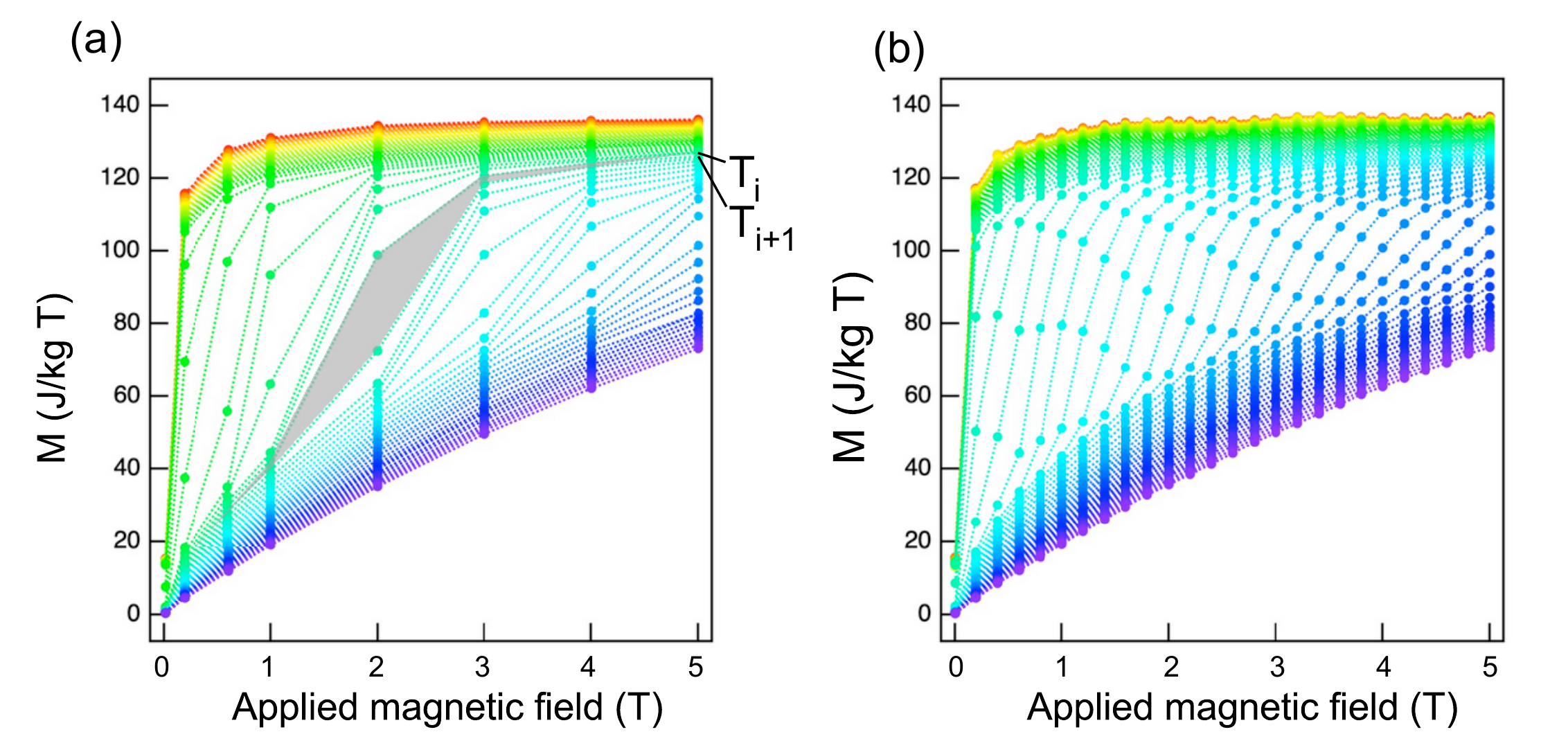}
  \caption{Schematic graph showing how linear interpolation along $H$ affects estimated $|\Delta$$S_{M}(T_{i},H)|$ value, in case of isothermal measurement. (a) Coarse magnetic field step (train data) case.  Difference in color stands for different measurement temperatures. Gray area corresponds to estimated $|\Delta$$S_{M}(T_{i},H)|$. (b) Fine magnetic field step case, simulated by neural networks model.}
  \label{fig:MH}
\end{figure}

\section{Operating system and installation guide}
The current proposed method have been tested only on Linux and MacOS (Intel) systems. We currently do not support Windows systems, however it might be possible to run using virtual machines, altough not recommended. In more details, the following OSes were tested by the authors:
Ubuntu LTS 20.04 and 18.04, CentOS 7, and MacOS 11 and 12 (both Intel).
A tutorial for how to get ready and perform the neural networks learning and simulation shown in this paper will be available at https://www.github.com/kensei-te/mat\_interp.

\section{Example of duration to perform neural networks learning}
Table SM1 shows the total duration it took for each computer to finish 30 trials of learning.  The used computers are the following: (i) Intel i9-9880H 2.3 GHz, (ii) Xeon Silver 4116 2.1 GHz, and (iii)Intel i7 10700K 3.8 GHz.  The number of parallel workers are set to be 5 for (i) and (iii), while it was 15 for (ii).  Since the result would be influenced by random seeds, we tried 5 times for each, and the averaged values are shown in the table.  Standard deviation of 5 times of 30 trials are also presented as numbers in parenthesis.  The number of train data points are 808, 2737, 767, 29977 for Section 3.1, 3.2, 3.3, 3.4, respectively.

%  Table cpus
\begin{table}[H]
   \begin{ruledtabular}
   \begin{tabular}{ccccc}
   
   Data & \makecell{Intel i9-9880H 2.3 GHz \\ 8 cores, MacOS11(Intel)} & \makecell{Xeon Silver 4116 2.1 GHz \\ 12x2 cores, CentOS7.6} & \makecell{Intel i7 10700K 3.8 GHz \\  8 cores, Ubuntu20.04} & \\\hline
   
   Section3.1 & 3m 30s (40s) & 3m 53s (29s) & 2m 36s (45s)  & \\\hline
   Section3.2 & 4m 30s (46s) & 3m 17s (26s) & 2m 31s (12s)  & \\\hline
   Section3.3 & 2m 53s (25s) & 3m 8s (30s)   & 2m 5s (35s)  & \\\hline
   Section3.4 & 13m 36s (2m 38s) & 6m 47s (47s) & 11m 14s (1m 46s)  & \\\hline
   
   \end{tabular}
   \end{ruledtabular}
   \caption{Comparison of ellapsed time required for PCs to finish learning through 30 trials of Baysian optimization for each dataset shown in the main text. The results are averaged values of 5 times of trials.  Numbers in parenthesis stand for standard deviation.}
   \end{table}

\begin{figure}[h]
\includegraphics[width=13 cm]{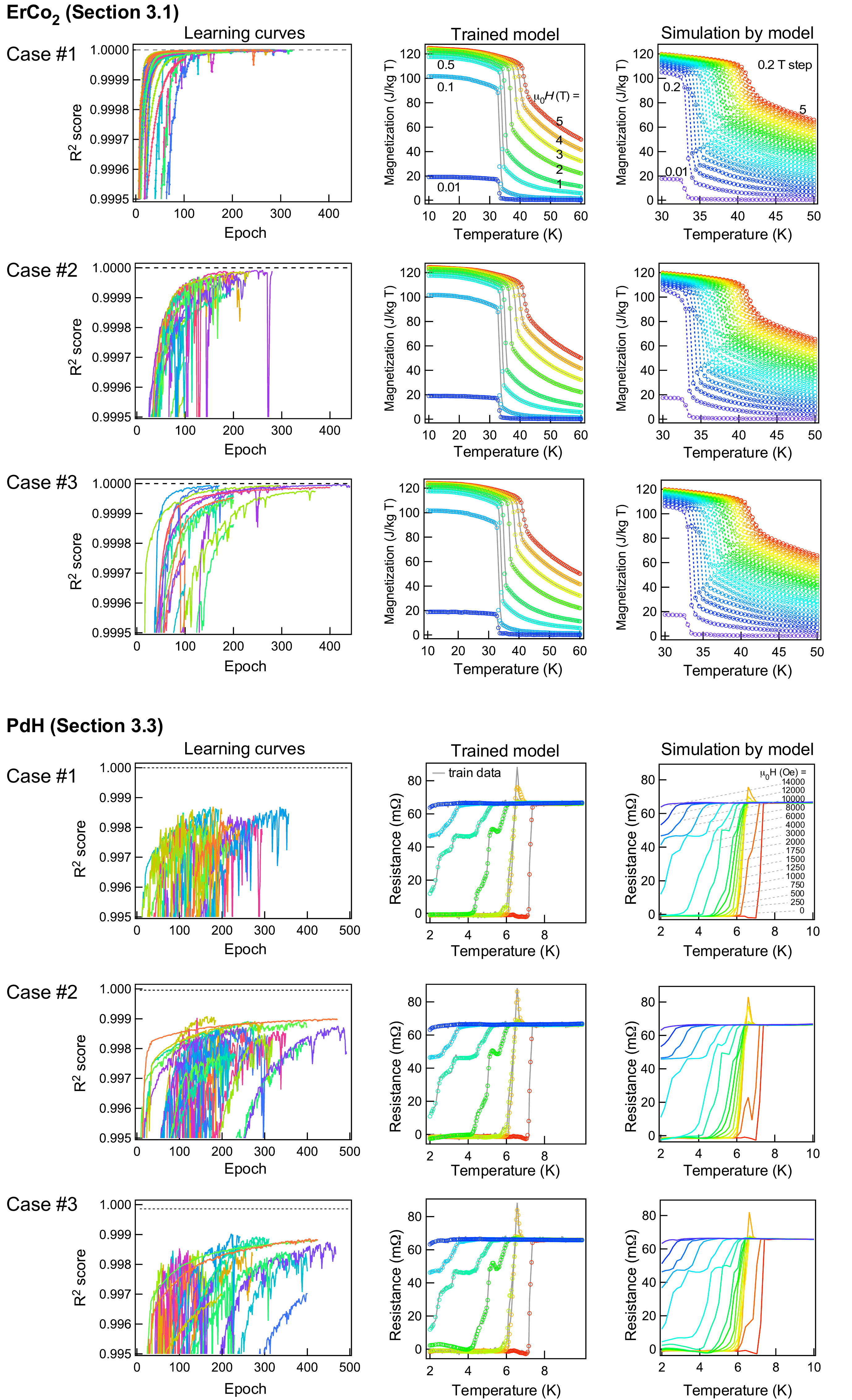}
\caption{Results of learning and simulations in individual cases of 30 trials for ErCo$_2$ (top) and PdH (bottom) data. For each material, case\#1 corresponds to the data in the main text.  Left: Learning curves. Colors represent different trials. Center: Simulation by learned models that acquired the best $R^2$ score in each case. Gray lines show train data.  Right: Simulation by models.}
\label{fig:stable}
\end{figure}

\section{Stability of learned and simulated results}
Here we show how much the learning curves and the resultant simulated value would differ among each 30 trials.  Two examples are shown, one for the data presented in Section 3.1 in the main text (Figure SM3), and another for the data presented in Section 3.3 (Figure SM4), where the latter is more influenced by noise.  Here each trial is allowed to keep running until it reachs a maximum epoch number of 500, unless it is stopped by either pruning or earlystopping.  After 30 trials of Bayesian optimization (initial 10 trials are random), both the $R^2$ score for training and simulated values show a convergence to some extent.

\section{Spectral lineshapes of ARPES image plot (Section3.4)}
Figure SM4 shows spectral lineshape of ARPES image plot in Section 3.4 of main text, where gray lines are train data and colored lines are simulated ones by neural networks model.  The spectra consist a number of peak structures and model can capture the structures as well as sensitivity of the machine such as relatively high intensity region around the edge.

\begin{figure}[h]
  \includegraphics[width=10 cm]{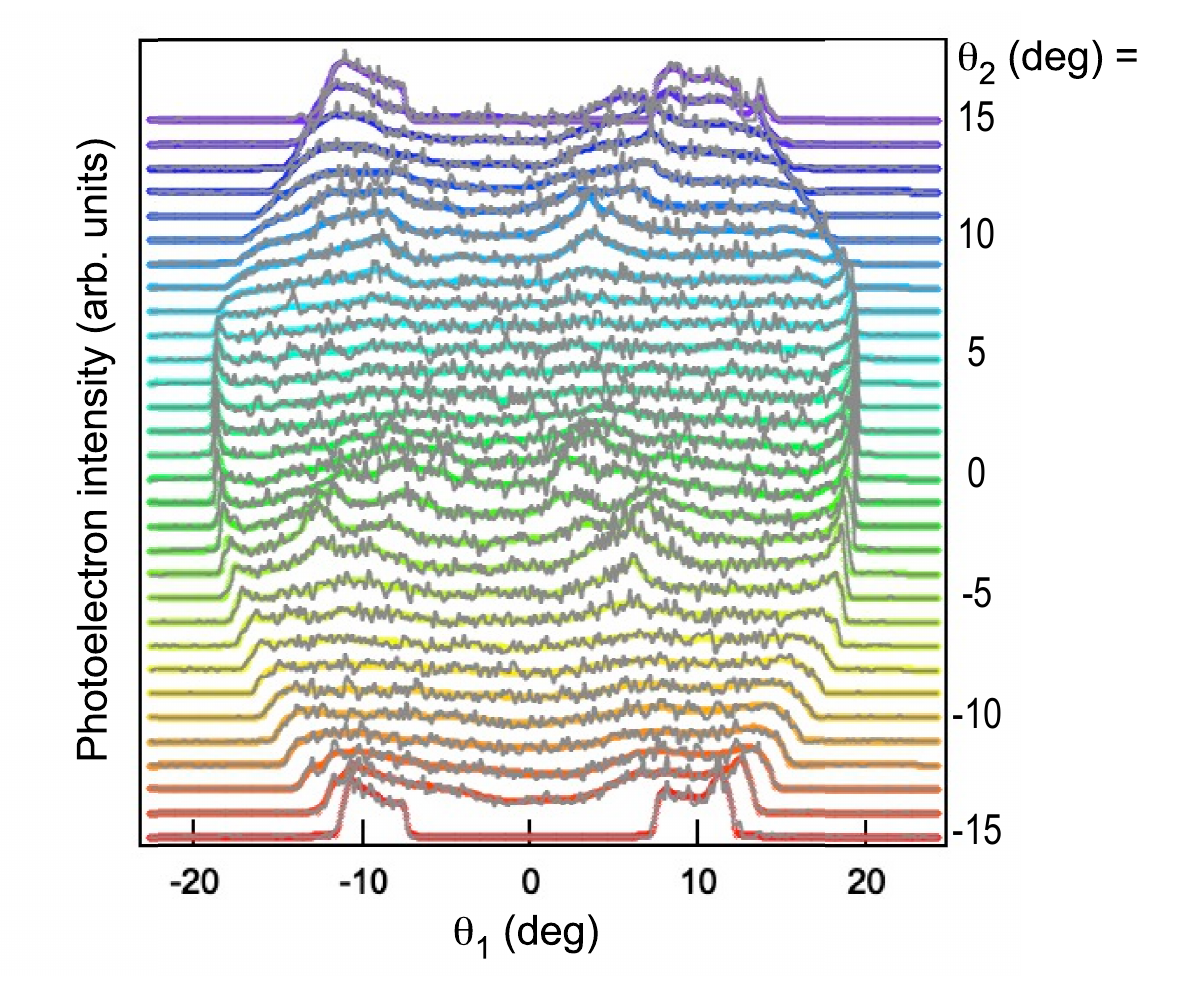}
  \caption{Spectral lineshapes of ARPES image plot, shown as Figure 5(a) and (b) in Section 3.4 of main text. Gray lines correspond to train data, while colored lines correspond to simulated ones by learned model.}
  \label{fig:arpes}
\end{figure}